\documentclass[aps,pra,twocolumn,notitlepage,superscriptaddress]{revtex4-1}
\usepackage{bm} \usepackage{graphicx} \usepackage{amsmath}
\usepackage{amsthm,stmaryrd}
\usepackage{amssymb} \usepackage{physics} \usepackage{txfonts}
\usepackage{color}
\usepackage{hyperref}
\usepackage[normalem]{ulem}

\newcommand{\NI}{N_{\mathrm I}}
\newcommand{\NP}{N_{\mathrm P}}
\newcommand{\NT}{N_{\mathrm T}}

\begin{document}

\title{Modelling Non-Markovian Quantum Processes with Recurrent Neural Networks }

\author{Leonardo Banchi }
\affiliation{ QOLS, Blackett Laboratory, Imperial College London, London SW7 2AZ, United Kingdom }

\author{Edward Grant}
\affiliation{Department of Computer Science, University College London, London 
WC1E 6EA, United Kingdom}

\author{Andrea Rocchetto}
\affiliation{Department of Computer Science, University College London, London 
WC1E 6EA, United Kingdom}
\affiliation{Department of Computer Science, University of Oxford, Oxford OX1 3QD, United Kingdom}

\author{Simone Severini}
\affiliation{Department of Computer Science, University College London, London 
WC1E 6EA, United Kingdom}

\begin{abstract}
	Quantum systems interacting with an unknown environment are notoriously difficult to model, 
	especially in presence of non-Markovian and non-perturbative effects. 
	Here we introduce a neural network based approach, which has the mathematical simplicity of the
	Gorini-Kossakowski-Sudarshan-Lindblad master equation, but is able to model non-Markovian 
	effects in different regimes. This is achieved by using recurrent neural networks 
	for defining Lindblad operators that can keep track of memory effects. 
	Building upon this framework, we also introduce a neural network architecture that is able to reproduce 
	the entire quantum evolution, given an initial state. 
	As an application we study how to train these models for quantum 
	process tomography, showing that recurrent neural networks are accurate over different 
	times and regimes. 
\end{abstract}

\maketitle

\section{Introduction}

Traditionally, in the physical sciences, the study of mathematical problems
for which no analytic solution is available involves modelling methods
leveraging a combination of approximation techniques (such as perturbation theory or 
semiclassical approaches) and the use of symmetries to reduce the
complexity of the problem. 
Recently, advances in machine learning 
\cite{Bishop2006,lecun2015deep}, have caused a surge in popularity of 
data-driven approaches, which instead rely on computational techniques that
exploit statistical correlations. 
Applications range from chaos theory~\cite{pathak2018model} to high energy physics~\cite{komiske2017deep}, 
eventually showing many applications and new perspectives
in the quantum domain
\cite{schuld2015introduction,biamonte2017quantum,ciliberto2018quantum,dunjko2018machine}. 
In particular, artificial
neural networks (a class of learning methods inspired by the functioning of
the brain) have been utilized in quantum many-body physics for ground state
estimation~\cite{carleo2017solving}, quantum state
tomography~\cite{torlai2018neural,yu2018reconstruction}, classification of phases of
matter~\cite{van2017learning}, entanglement estimation \cite{gray2017measuring}, 
and to identify phase transitions~\cite{carrasquilla2017machine}. Although the
theoretical understanding of the effectiveness of these models is currently
limited, some recent papers have established connections between
neural networks and more standard frameworks 
such as renormalization group~\cite{mehta2014exact}, tensor
networks~\cite{levine2017deep,levine2018bridging,chen2018equivalence}, and
complexity theoretic tools~\cite{rocchetto2017learning}.
Moreover, classical optimization techniques borrowed from supervised machine 
learning have been employed to optimize the dynamics of
many-body systems 
\cite{banchi2016quantum,innocenti2018supervised,eloie,las2016genetic,zahedinejad2016designing}
and parametric quantum  circuits 
\cite{verdon2018universal,grant2018hierarchical,melnikov2018active,huggins2018towards,benedetti2018adversarial,bukov2017reinforcement,niu2018universal}.


Open quantum systems 
\cite{breuer2002theory,rivas2012open} present further challenges. Here, 
any modelling effort must take into account that the system interacts with a surrounding 
environment, whose microscopic details are usually unknown.  The resulting effects can only 
be treated phenomenologically and significantly increase the complexity of the model, 
especially in the non-Markovian regime. In this regime, 
exact non-perturbative master equations~\cite{ishizaki2009unified} or 
quantum maps~\cite{pollock2018non} operate 
by entangling the system with  ancillary degrees of freedom whose dimension grows 
with time. This makes exact simulations extremely challenging even for 
low dimensional Hilbert spaces. Therefore, in larger systems, non-Markovian effects 
can only be modelled in an approximate fashion, e.g. by neglecting quantum correlations 
or by assuming weak couplings between system and environment. When these 
approximations are justified, phenomenological models with 
a reduced number of parameters are normally accurate. However, these 
approximations are  often  violated, e.g., in quantum biology \cite{ishizaki2009unified},
so it is important to study alternative mathematical structures.

Machine learning methods offer new ideas for modelling non-Markovian effects when standard assumptions do not hold. The intuition behind these approaches is that, if there exists an efficient description of the system, this can be learnt from data without using explicit modelling assumptions. In particular, neural networks based learning techniques have shown that it is
possible to learn complex functional dependencies in time series directly from 
the data, without trying to make theoretical assumptions that may be unjustified 
(such as the weak coupling limit) or hard to derive from phenomenological observations.
In order for these methods to succeed, it is crucial to define the correct neural
network structure that can quickly learn the underlying rule to reproduce 
the unknown functional form displayed in the data.

Here we investigate the ability of neural networks to model the non-Markovian 
evolution of open quantum systems, using a fixed number of parameters that does not 
grow with the number of time steps. We focus on Recurrent Neural Networks (RNNs),
a type of artificial neural network specifically designed to model dynamical
systems with possibly long range temporal correlations. In the quantum setting,
RNNs have been previously employed for quantum control \cite{august2017using,ostaszewski2018approximation}.  We consider two main 
applications. In the first one, we define a master equation which has the 
mathematical simplicity of the Gorini-Kossakowski-Sudarshan-Lindblad 
master equation \cite{gorini1976completely,lindblad1976generators}, but 
is nonetheless able to model non-Markovian effects via the memory cells 
included in RNNs. In the second application, we define an RNN able to reproduce 
the entire quantum evolution given an initial state, without introducing any 
master equation. Our RNN-based frameworks share some 
similarity with collisional models 
\cite{ciccarello2013collision,ziman2002diluting,giovannetti2012master,rybar2012simulation,mccloskey2014non,ciccarello2017collision,campbell2018system} or $\epsilon$-machines \cite{binder2018practical}, 
where explicit memory effects are introduced by using ancillary quantum systems. 
However, there are notable differences. One of the advantages of RNNs is in their ability to 
learn directly from data how to compress complex time series with a few 
constant parameters. 
Moreover, there are different RNN architectures and it is possible to select 
the most appropriate one to model the expected long-range memory in the 
time evolution. 

We will show that RNNs provide a convenient mathematical framework 
to define the so-called memory kernel \cite{breuer2002theory}, such that the 
resulting time-local master equation has a completely positive solution. 
The reconstruction of the memory kernel from quantum state sequences, 
namely from quantum process tomography, 
has been the subject of different studies in recent years 
\cite{cerrillo2014non,pollock2017tomographically,pollock2018non,howard2006quantum,yuen2011quantum,bellomo2010tomographic}.
Most approaches considered in the literature start with some assumptions on the 
microscopic model, and then fits the free parameters given the experimental data. 
These assumptions are required to avoid having an exponentially  growing number of 
free parameters for increasing number of time steps. 
However, these assumptions are normally 
uncontrollable, especially when the details of the microscopic model are 
unknown.  Our main result is to show that RNN-based quantum processes and master equations are able to learn efficient representations of non-Markovian quantum evolutions directly from data sequences and without making any assumption on the underlying physical model.
Indeed, even the 
Hamiltonian of the system can be learnt during this process. Note that,
while the Hamiltonian can be learnt with alternative methods, e.g. Bayesian 
inference \cite{wiebe2014hamiltonian}, the memory kernel is much more difficult, 
as its functional form, beyond the perturbative regime,  is unknown. 
The main use of RNNs is 
therefore as a ``compression'' method, to effectively model 
complex quantum correlations between system and environment with a constant and 
fixed number of parameters. 
As a relevant application for our technique, we 
consider quantum process tomography and show that RNNs are able to model 
the physical evolution of quantum systems over different regimes. 

The paper is structured as follows. In Sec.~\ref{s:background} we 
present relevant technical background. Specifically, in Sec.~\ref{s:backgroundNM} 
we introduce non-Markovian quantum processes and common master equations 
to describe them. In Sec.~\ref{s:RNN} we introduce the 
RNN architecture employed in this paper. The main ideas are presented in 
Sec.~\ref{s:idea}, where we introduce our RNN-based master equation (Sec.~\ref{s:ideaME})
and RNN-based quantum process (Sec.~\ref{s:qp}), with applications for 
process tomography (Sec.~\ref{s:envlearning}). Numerical experiments are 
presented in Sec.~\ref{s:numerics} and conclusions are drawn in Sec.~\ref{s:conclusions}.

\section{Background} \label{s:background}
\subsection{Non-Markovian processes} \label{s:backgroundNM}
Quantum systems, even the purest ones, are inevitably in contact with an environment. 
Because of this normally unknown interaction, quantum evolution deviates from 
the predictions of the Schr\"odinger equation, and different extensions have been 
proposed \cite{breuer2002theory,breuer2016colloquium}. 
When the interactions inside the environment 
happen on timescales much shorter than the internal timescales of the system, 
then a Markovian approximation is usually appropriate, and the evolution can be 
modeled via the Gorini-Kossakowski-Sudarshan-Lindblad (GKSL) master equation
\cite{gorini1976completely,lindblad1976generators}
\begin{align}
	\label{e:GKSL}
	\frac{\partial \rho(t)}{\partial t} = & \mathcal L[\rho(t)]~, \\ \nonumber
	\mathcal L[\rho] = &{-}i[H,\rho] + \sum_\mu \left[
		L_\mu\rho L_\mu^\dagger -\frac12\left\{L_\mu^\dagger L_\mu,\rho\right\} 
	\right]~,
\end{align}
where $H$ is the Hamiltonian, which models the noise-free case, and 
$L_\mu$ are called Lindblad operators. The superoperator $\mathcal L$ is 
called Liouvillian. Mathematical properties of the above equation are well
understood \cite{rivas2012open}. For any choice of $H$ and $L_\mu$ the solution of the 
master equation $\mathcal E_t= e^{\mathcal Lt}$ defines a completely positive 
trace preserving linear map, and is thus a mathematically well-posed 
mapping between states to states.  With properly chosen Lindblad operators, 
the above master equation
models the most general Markovian evolution
\cite{gorini1976completely,lindblad1976generators}. 
However, the Markovian approximation is not accurate in many situations, 
for instance when the interactions inside the environment have comparable strengths to the interactions inside the system 
\cite{ishizaki2009unified,banchi2013analytical}. 
In that case the master equation has to be modified to take into account 
non-Markovian effects. One of the first and most accurate descriptions 
of non-Markovian evolution is the Nakajima-Zwanzig (NZ) master equation 
\cite{breuer2002theory}
\begin{align}
	\frac{\partial \rho(t)}{\partial t} = {-}i[H,\rho(t)] + 
	\int_0^t ds\, \mathcal K^{\rm NZ}_{t-s}[\rho(s)] + \mathcal I(t)~,
	\label{e:NZ}
\end{align}
where $\mathcal K^{NZ}_{t}$ is a super-operator, the so-called 
{\it memory kernel}, which describes the interaction 
with the environment, while $\mathcal I(t)$ is due to the initial correlations 
between system and environment. If system and environment are initially 
uncorrelated, then $\mathcal I(t)=0$ for all $t$. It is clear that the 
above equation describes non-Markovian processes, because the state at time $t+dt$ 
depends not only on $\rho(t)$ but also on the states $\rho(s)$ for $s<t$. 
The Nakajima-Zwanzig equation is at the basis of powerful Green function 
methods to study the spectral properties of the system 
\cite{mukamel1999principles,zhang2012general,jang2003theory,banchi2013analytical},
since the convolution disappears in the frequency domain.
On the other hand, in the time domain 
the above equation is not easy to solve numerically. 
To avoid this problem, a different but 
equally accurate master equation has been proposed, the 
so-called time-convolutionless (TCL) 
master equation \cite{breuer2002theory}, which reads
\begin{align}
	\frac{\partial \rho(t)}{\partial t} = {-}i[H,\rho(t)] + 
	\int_0^t ds\, \mathcal K^{\rm TCL}_{t-s}[\rho(t)] + \mathcal I(t)~.
	\label{e:TCL}
\end{align}
The main formal difference between Eq.~\eqref{e:TCL} and 
Eq.~\eqref{e:NZ} is that the whole history of states is fed into 
the NZ  master equation, while in the TCL case the master equation explicitly 
depends on $\rho(t)$ only, and all non-Markovian effects are included into the 
memory kernel.  As such, the non-Markovian nature of the process resulting from 
Eq.~\eqref{e:TCL} is less obvious, but it is known that both NZ and TCL 
master equations can describe the same processes. Indeed, there are formal 
mappings between $\mathcal K^{\rm NZ}$ and $\mathcal K^{\rm TCL}$ 
such that both Eq.~\eqref{e:TCL} and Eq.~\eqref{e:NZ} produce the 
same physical evolution \cite{chruscinski2010non,breuer2002theory}.

Although TCL master equations are relatively easy to solve numerically, 
the main problem is that the interaction with the environment is normally 
unknown. In other terms, while $H$ is typically well characterized experimentally, the memory 
kernel depends on environmental quantities such as the temperature, but also 
on the spectral properties of the environment which are normally unknown. 
Non-linear spectroscopy can be used to find the spectral density 
\cite{mukamel1999principles}, but the latter does not completely characterizes the memory 
kernel, without introducing further assumptions. The most commonly employed approximation is  
the assumption of 
a weak coupling between system and environment, such that the memory kernel can be formally
obtained using perturbation theory. However, there are cases where these approximations are not 
justified,  e.g. in quantum biology \cite{ishizaki2009unified},
where the strength of the interaction with the environment is comparable with the internal interactions inside 
the system. To go beyond perturbation theory, we introduce then a RNN-based 
non-Markovian quantum process, where memory effects can be learnt directly 
from data.

\subsection{Recurrent Neural Networks}\label{s:RNN}

RNNs are a class of neural networks designed to model data sequences like time series 
\cite{schmidhuber2015deep}. 
To understand their functioning, it is helpful  
to compare them with more standard feedforward networks. In feedfordward neural networks 
the input data $s^0$ propagates throughout many intermediate (hidden) layers before 
reaching the final output layer. Here ``propagate'' means that, step by step, 
the state $s^{\ell+1}$ of the $(\ell{+}1)$-th layer is updated, given the state of $\ell$-th layer, 
as $s^{\ell+1} = f(W^{\ell} s^{\ell} + w^\ell)$ where $W^{\ell}$ is a weight matrix, $w^{\ell}$
a weight vector and $f$ is some non-linear function. The state at the final layer 
of the network (output layer) depends on all the weight matrices and vectors. 
Training is performed by updating those weights such that the neural network 
learns some desired input-output relationship hidden in the data. 

In case of temporal data, each input has 
also an explicit time dependence $s^0_t$. Although, in principle, one could 
still use a giant feedforward network with these data, this is rarely the 
optimal choice, because the number of free parameters quickly increases with the number of 
time steps. RNNs solve this issue with a 
more advanced architecture which 
is tailored for temporal data. In RNNs, the update rule for the
hidden layers at time $t$ not only depends on the states $s^\ell_t$, but 
also on the states at previous times. In other terms, the update rule is 
$s^{\ell+1}_t = f(W^{\ell} s^{\ell}_t + w^\ell, s^{\ell+1}_{t{-}1})$, where 
$s^0_t$ is the input temporal sequence. The free parameters $W^\ell$ and $w^\ell$ 
do not depend on $t$, and memory of the past is taken into account by the function 
$f$, which compresses and saves relevant informations of previous sequences into memory cells. 
This architecture allows RNNs to learn temporal sequences using a relatively 
small number of parameters, even when the temporal data has long-range memory effects. 
The mapping between $s^{\ell}_t$ and 
$s^{\ell+1}_t$ defines a RNN cell. 

\begin{figure}[t]
	\centering
	\includegraphics[width=\linewidth]{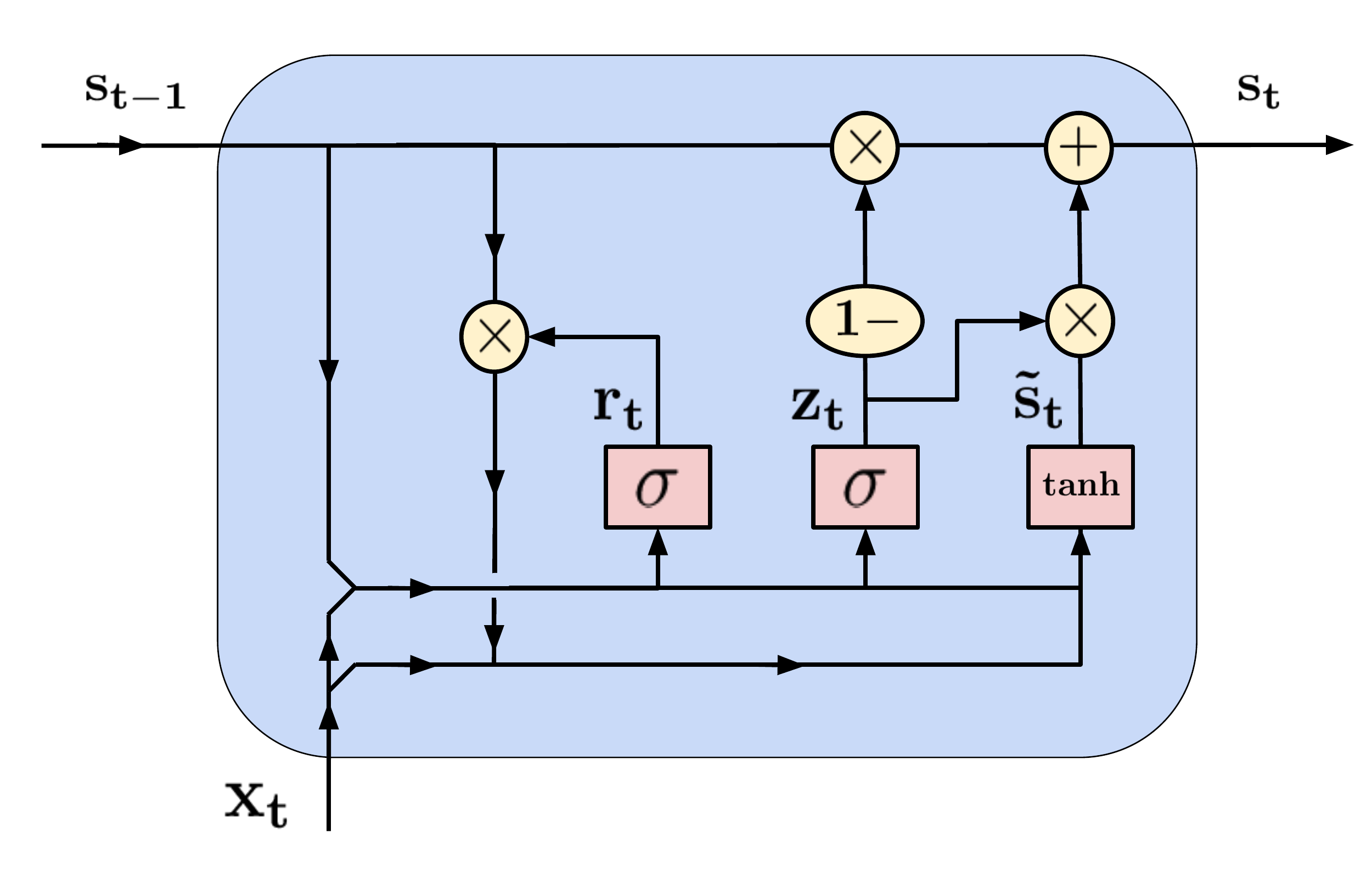}
	\caption{
		{\bf Network diagram of the GRU cell }. 
		The output $s_t$ and input $s_{t-1}$ represent the state at times $t$ and $t-1$, respectively, while 
		$x_t$ is an auxiliary input that depends on the previous times, before $t-1$. 
		Rectangles represent neural network layers. Circles represent entry-wise operations. Bifurcations represent copy operations and joined lines represent concatenation. 
		Details are presented in Appendix~\ref{a:GRU}.
	} 
	\label{fig:GRU}
\end{figure}

In this work we use a variant of RNN called Gated Recurrent Unit (GRU), whose basic cell is shown in Fig.~\ref{fig:GRU}  and discussed  in Appendix~\ref{a:GRU}. GRUs use a gating mechanism that allows them to better model long-term dependencies than more simple RNNs~\cite{cho2014learning}. GRUs are based on a type of RNN cell called Long Short-Term Memory (LSTM), but can be more efficient than LSTMs for comparable performance~\cite{hochreiter1997long,chung2014empirical}. GRU and LSTM are commonly used and achieve state of the art performance for sequence modelling across multiple domains, including machine translation, image captioning and forecasting~\cite{lipton2015critical}.

The GRU state $s_{t}$ is a linear interpolation of the previous state $s_{t-1}$ and a candidate state $\tilde{s}_{t}$, which depends on the auxiliary input  $x_{t}$. The input $x^{j}_{t}$ for a depth $j$ cell at time $t$ is the state from the cell in the previous layer $s^{j-1}_{t}$. GRU cells can be stacked to form a deep GRU network. More details can be found in Appendix~\ref{a:GRU}.

\section{Main idea}\label{s:idea}

In this section we present the three main contributions of this paper which all leverage on the modelling capabilities of RNNs to describe non-Markovian dynamics of open quantum systems. First, we describe a master equation approach. Second, we use RNNs to predict the time evolution of quantum state under a non-Markovian quantum process. Third, we show how these two techniques can be utilized to perform quantum process tomography.

\subsection{RNN Quantum Master Equation} \label{s:ideaME}

\begin{figure*}[t]
    \centering
    \includegraphics[width=1.0\textwidth]{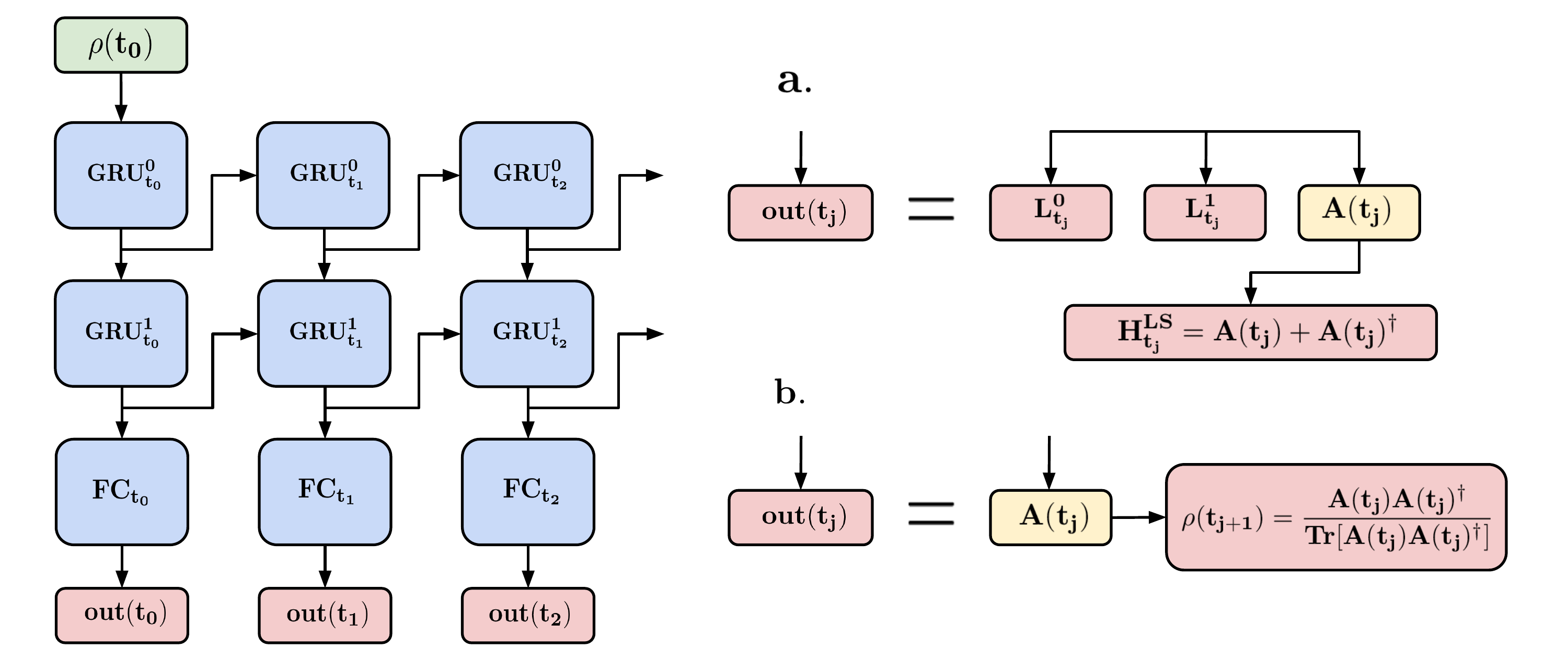}
		\caption{\textbf{RNN architectures for open quantum systems}: schematic of a one-to-many deep RNN used to approximate the master equation and to model the non-Markovian dynamics. Both networks take as input $\rho(t_{0})$ (green cell) and comprise of two GRU layers and a fully connected (FC) layer (blue cells). The yellow cells show the initial network output $A(t_j)$ before the post-processing that makes $H_{t_j} ^{LS}$ Hermitian and $\rho(t_{j+1})$ a valid density operator. Panel \textbf{a.} shows the output for the QML master equation, i.e. the predicted value of $H^{LS}_{t}$ and $L^{\mu}_{t}$ at intervals $\Delta{t}$ (red cells). Panel \textbf{b.} shows the output for a RNN modelling the time evolution of a quantum state undergoing a non-Markovian quantum process, i.e.  the predicted value of $\rho(t)$ at intervals $\Delta{t}$ (red cells).}
    \label{fig:GRU_net}
\end{figure*}

We postulate a quantum evolution similar to Eq.~\eqref{e:GKSL}, namely 
\begin{equation}
	\label{e:GKSLRNN}
	\frac{\partial \rho(t)}{\partial t} =  \mathcal L_{{\le}{t}}[\rho(t)]~,
\end{equation}
where the notation ${\le}t$ refers to a superoperator that not only depends on 
$t$, but also on the entire history before time $t$, as in the TCL master equation
\eqref{e:TCL}. 
A convenient choice is then that of the GKSL form
\begin{align}
	\label{e:QRNME}
	\mathcal L_{{\le}t}&[\rho] = {-}i[H + H^{LS}_{{\le}t},\rho] + \cr 
													 & + \sum_\mu \left[
		L^\mu_{{\le}t}\rho L_{{\le}t}^{\mu\dagger} -
		\frac12\left\{L_{{\le}t}^{\mu\dagger} L_{{\le}t}^\mu,\rho\right\} 
	\right]~,
\end{align}
where $H^{LS}$ is a ``Lamb-shift'' term, namely a correction to the Hamiltonian 
induced by the environment, while $L^\mu$ are Lindblad operators. The reason for this 
choice is that for small enough $\Delta_t$ the time evolution is simply given 
by $\rho(t+\Delta_t) \approx e^{\Delta_t \mathcal L_{{\le}{t}}}[\rho(t)]~$,
and since $\mathcal L_{{\le}t}$ is in the GKSL form, 
$e^{\Delta_t \mathcal L_{{\le}t}}$ is a completely positive trace preserving quantum channel, 
mapping states to states. If $H^{LS}$ and $L^\mu$ are simply time-dependent functions, 
namely they only depend on $t$ and not on previous times, then the dynamics generated by
above master equation is always Markovian \cite{chruscinski2010non}. The main 
idea of this work is to use a RNN to define each Lindblad operator $L_{{\le}t}^\mu$ and the correction Hamiltonian $ H^{LS}_{{\le}t_j}$, see fig.~\ref{fig:GRU_net}(a). In order to ensure the Hermiticity of the $ H^{LS}_{{\le}t}$ operator, we construct it as $H^{LS}_{{\le}t} = A(t_j) + A(t_j)^{\dagger}$, where $A(t_j)$ is the output of the network and $A(t_j)^{\dagger}$ its conjugate transpose.
Since in RNNs the predicted output at time $t$ 
depends on the entire history at previous times, this 
parametrization is expected to accurately reproduce genuinely non-Markovian effects, 
even with possible long-range dynamical correlations. The master equation then 
resembles a TCL master equation \eqref{e:TCL}, but 
where the complicated memory superoperator is expressed via a simpler GKSL form with 
 RNNs. We call our Eq.~\eqref{e:QRNME} the Quantum Recurrent Neural 
 (QRN) Master Equation. Similarly, we call the operators  $L_{{\le}t}^\mu$ recurrent Lindblad operators. A schematics of the resulting neural network is shown in Fig.~\ref{fig:GRU_net}(a). 

\subsection{RNN Quantum Processes} \label{s:qp}
For any initial time $t_0$ the mapping between the initial state $\rho(t_0)$ and the state
at time $t$, namely 
\begin{equation}
	\rho(t) = \mathcal E(t,t_0)[\rho(t_0)]~,
\end{equation}
defines a completely positive map, assuming no initial correlation with the environment. 
For any intermediate times $t_0<\tau<t$ the mapping
$\mathcal E(\tau,t_0)$ is always completely positive. When also 
$\mathcal E'(t,\tau):= \mathcal E(t,t_0) \mathcal E(\tau,t_0)^{-1}$ 
is completely positive, then the mapping is called {\it divisible} 
\cite{vacchini2011markovianity}. Divisibility is another way of characterising 
Markovianity, as quantum processes obtained from the GKSL master equation are always divisible 
with $\mathcal E(t',t) = \mathcal T \exp\left(\int_t^{t'} \mathcal L_s\, ds\right)$.

In the previous section we have introduced the QRN master equation and shown that it 
can model non-Markovian effects, even when the evolution between intermediate steps 
is completely positive. This is possible because the RNN keeps a compressed record 
of the previous evolution. Using the formalism of the previous section we can indeed
write 
\begin{equation}
	\mathcal E(t,t_0) = \overleftarrow{\prod_j}  \mathcal E^{QRN}(t_j+ \Delta_t,t_j) 
	+ \mathcal O(\Delta_t^{-2}),
	\label{e:RNNQP0}
\end{equation}
where $\overleftarrow{\prod}_j X_j$ is the ordered product $\cdots X_3 X_2 X_1$. 
Moreover,
\begin{equation}
	\mathcal E^{QRN}(t_j+ \Delta_t,t_j) = e^{\Delta_t \mathcal L_{{\le}{t_j}}}
\end{equation}
is completely positive, being the operator exponential of a Liouvillian, but 
depends on the previous evolution, via the recurrent Lindblad operators. As such, 
the total map \eqref{e:RNNQP0} is not divisible. 

Based on this analogy, we can now drop the master equation
and define a non-Markovian process as 
\begin{equation}
	\mathcal E(t,t_0) = \overleftarrow{\prod_j}  \mathcal E^{RNN}_{{\le}t_j}(t_{j+1},t_j) ~,
	\label{e:RNNQP}
\end{equation}
where $\mathcal E^{RNN}_{{\le}t_j}(t_{j+1},t_j)$ is completely positive, but depends on 
the entire history before $t_j$. Each $\mathcal E^{RNN}_{{\le}t_j}(t_{j+1},t_j)$, being 
completely positive, outputs 
a valid quantum state at intermediate times $\rho(t_j)$ and, being a RNN, 
then updates its internal memory. 
Complete positivity can be ensured by using the Kraus decomposition or the 
environmental representation. 

For better comparison with the QRN master equation, in this work 
we use the simpler strategy shown in Fig.~\ref{fig:GRU_net}(b), 
where we use the output $A(t_j)$ of the network to define a 
density operator via 
$\rho(t_{j+1}) = A(t_j)A(t_j)^\dagger / \operatorname{Tr}[A(t_j)A(t_j)^\dagger]$.  
This ensures that the states $\rho(t_j)$ are 
valid density operators, throughout the entire evolution.

\subsection{Application: Quantum Process Tomography }\label{s:envlearning}

In this work we apply our QRN master equations and RNN quantum processes for quantum process tomography. We consider the following setup. 
We assume that the quantum system can be 
initialized in different initial states $\rho_\alpha(0)$ where 
$\alpha=1,\dots,\NI$, for some number $\NI$. For each initialization, we assume that 
it is possible to perform full-quantum tomography at some time steps 
$0\le t_j \le T$ for $j=1,\dots,\NT$ and some $\NT$. After this procedure we 
are able to reconstruct the time evolutions $\rho_\alpha(t_j)$ for
different times and initializations. Here we consider uniformly separated times 
where $t_j = j \Delta_T $ and $\Delta_T = T/\NT$, though it is straightforward 
to generalize this procedure to non-uniform sequences. 
Each state tomography requires $\mathcal O(d^2)$ measurements, where 
$d$ is the dimension of the system's Hilbert space, so
the total cost of reconstructing these sequences is $\mathcal O(d^2 \NT \NI)$. 
Once these sequences are obtained, we use them to train the neural network. 

We consider two cases. In the first one we train a RNN to learn the quantum state evolution,
as shown in Fig.~\ref{fig:GRU_net}(b). Here we assume no knowledge about the system's Hamiltonian
or interaction with the environment. Training is then performed by minimising a cost 
function 
\begin{equation}
	J_p = \frac1{\NI \NT} \sum_{\alpha=1}^{\NI} \sum_{j=1}^{\NT} 
	\| \rho_\alpha(t_j)- \tilde{\rho}_\alpha(t_j)\|~,
	\label{e:JP}
\end{equation}
where $\rho_\alpha(t_j)$ are the training data,
$\tilde{\rho}_\alpha(t_j)$ are the states outputted by the RNN, and $\|\cdot\|$ is any operator norm. 

In the second application we aim at reconstructing  $H^{LS}_{{\le}t}$ 
and the recurrent Liouvillian operators $L^\mu_{{\le}t}$ 
entering in the QRN master equation, where, on the other hand, 
we assume that the Hamiltonian $H$ is known. 
The latter assumption can always be relaxed, as Hamiltonian evolution can 
be fully included into the correction Hamiltonian $H^{LS}$,  
which is learnt from the data. 
In the QRN master equation, the operators $H^{LS}_{{\le}{t}}$ and $L^{\mu}_{{\le}{t}}$ 
are obtained from the output of the RNN, as shown in Fig.~\ref{fig:GRU_net}(b). 
To train the RNNs to predict $H^{LS}$ and $L^{\mu}$
given a starting state $\rho_{\alpha}(t=0)$  we propose the use 
of the differential equation \eqref{e:GKSLRNN} to define a cost function. 
For classical time series, a similar approach \cite{schmidt2009distilling} 
has been employed for extracting natural laws from experimental data. 

To explain our idea, let us first consider the opposite scenario, where
the differential equation \eqref{e:GKSLRNN}, with all of its operators, is already known,
while the states $\rho(t_j)$ are not. In this common case, 
the states $\rho(t_j)$ at different times are evaluated from the 
numerical integration of the master equation. The latter can be obtained 
with an $n$-order Runge-Kutta integrator \cite{suli2003introduction}  which,
in general, can be formally  written as 
\begin{equation}
	\rho_\alpha(t_{j+1}) = 	\rho_\alpha(t_{j}) + \Delta_T 
\mathcal L^{{\rm RK},n}_{{\le}t_j}[\rho_\alpha(t_j)]~,
\end{equation}
where $\mathcal L^{{\rm RK},n}_{{\le}t}$ is $n$-th order Runge-Kutta integration step, 
which can be explicitly obtained for any $n$ (see e.g. Ref.~\cite{suli2003introduction}). 
For instance, to the first order 
$\mathcal L^{{\rm RK},1}_{{\le}t}$ is simply $\mathcal L_{{\le}t}$. 
To summarize, when $H^{LS}_{{\le}{t}}$ and $L^{\mu}_{{\le}{t}}$ are known, we can use 
a Runge-Kutta integrator to obtain the time evolution $\rho(t_j)$.

We now consider the opposite problem, namely where many time sequences 
$\rho_\alpha(t_j)$ are already known, while the operators 
$H^{LS}_{{\le}{t}}$ and $L^{\mu}_{{\le}{t}}$ in the QRN master equation 
are not. This is like assuming that the solutions of a differential equation
are known and from them we want to reconstruct the differential equation itself. 
Based on the analogy with numerical integration via Runge-Kutta, we propose 
to use the following cost function:
\begin{equation}
	J = \frac1{\NI \NT} \sum_{\alpha=1}^{\NI} \sum_{j=1}^{\NT} 
	\left\|\rho_\alpha(t_{j+1}) - 	\rho_\alpha(t_{j}) - \Delta_T 
\mathcal L^{{\rm RK},n}_{{\le}t_j}[\rho_\alpha(t_j)]\right\|^{2}_F ~,
\label{e:cost}
\end{equation}
where $\|\cdot\|_F$ is the Frobenius norm. 
The intuitive idea behind the above cost function is that of making the measured data sequences 
as close as possible to those coming from the numerical solution of a master equation. Minimizing
the cost function is then equivalent to finding the best QRN master equation compatible with 
the measured sequences $\rho_\alpha(t_j)$.
It is expected that a higher order integrator (large $n$) performs better, especially 
for larger $\Delta_T$, but requires 
heavier numerical computations.

\section{Numerical experiments}\label{s:numerics}
\subsection{Learning quantum state sequences} \label{s:statelearn}
We mimic experimental data by numerically generating sequences of states, which are then
used to train the neural network. 
To generate the training data, we consider a simple yet important model of 
spontaneous decay of a two-level system \cite{breuer1999stochastic,breuer2016colloquium},
described by the master equation 
\begin{align}
  \label{e:2lev}
	\frac{\partial}{\partial t}&\rho(t) =-i 
  [\omega \sigma_z,\rho(t)] + \\
	&+\gamma(t)\left( \sigma^-\rho(t)\sigma^+
	-\frac{1}{2}\left\{\sigma^+\sigma^-,\rho(t)\right\}
\right),\nonumber
\end{align}
where $\{x,y\}=xy+yx$, $\sigma^{\alpha}$ for $\alpha=x,y,z$ are the Pauli matrices, 
$\sigma^\pm=(\sigma^x\pm i\sigma^y)/2$, $\omega$ is the Rabi frequency of 
oscillations around the $z$ axis. The parameter
\begin{equation}
\label{e:rabi}
  \gamma(t)=\frac{2\gamma_0\lambda \sinh({\eta t/2})}
  {\eta \cosh({\eta t/2})+\lambda\sinh({\eta t/2})}.
\end{equation}
is the decay rate 
with $\eta =\sqrt{\lambda^2-2\gamma_0\lambda}$. When $\lambda>2\gamma_0$ the function 
$\gamma(t)$ is always positive, so Eq.~\eqref{e:2lev} takes the GKSL form \eqref{e:GKSL}
and, as such, defines a Markovian evolution. On the other hand, for $\lambda<2\gamma_0$ 
the function $\gamma(t)$ can be negative and the dynamics displays non-Markovian effects
\cite{breuer2016colloquium,chruscinski2014degree,bylicka2014non}.
The above two level system can also model a excitation energy transfer in 
biological dimers \cite{iles2016energy}. 

\begin{figure}[t]
    \centering
    \includegraphics[width=.49\textwidth]{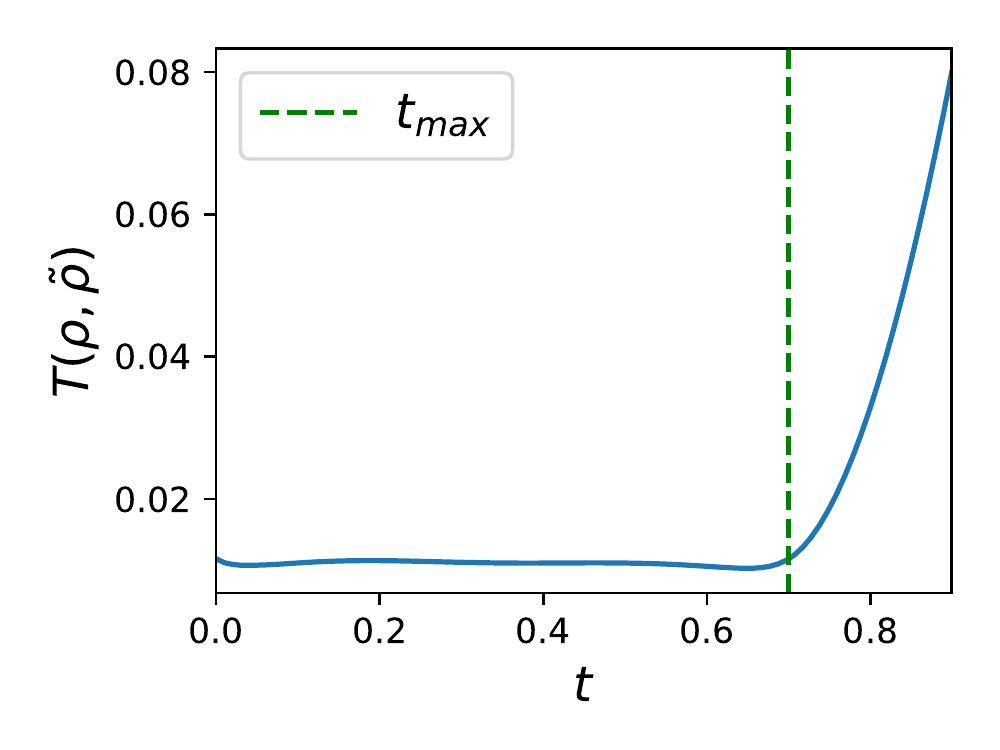}
    \caption{\textbf{(EXP1) Learning the evolution of a noisy two level system:} 
			the mean trace distance $T(\rho(t),\tilde\rho(t)) = \frac1\NP\sum_\beta D(\rho_\beta(t),\tilde\rho_\beta(t))$
			between true states $\rho_\beta(t_j)$ and predicted states $\tilde{\rho}_\beta(t_j)$ as a function of time. 
			The time $t_{max}$ is the maximum time used during training.
		All the experiments run with the following parameters $\omega = 1$ for
		Eq.~\ref{e:2lev} and $\lambda = 2$, $\gamma_0 = 0.5$ for Eq.~\ref{e:rabi},
		and a discretization interval of $\Delta t = 0.01$. The simulations run
		over $3000$ training examples and predictions were tested over $1000$ test
		examples. All the evolutions run for a time of $t_{max} = 0.7$. 
    }
    \label{fig:trace_distance}
\end{figure}

We use the above model to obtain training data for the RNN architecture shown in 
Fig.~\ref{fig:GRU_net}(b). Each training sequence 
$\rho_\alpha(t_j)$, for $\alpha=1,\dots,\NI$, has been obtained by first 
choosing a random initial state $\rho_\alpha(0)$ and then solving the master equation 
Eq.~\eqref{e:2lev} to get the states  $\rho_\alpha(t_j)$ at subsequent times,
up to a maximal time $t_{max}$. 
These data sequences were then used to train a GRU neural network, by minimising 
the cost function \eqref{e:JP}. 
After training, we test the accuracy of the 
neural network by generating a new sequence of states $\rho_\beta(t_j)$, and the 
RNN prediction $\tilde\rho_\beta(t_j)$, for $\beta=1,\dots,\NP$. As before,
$\rho_\beta(t_j)$ is obtained by selecting 
a random initial state $\rho_\beta(0)$ and solving Eq.~\eqref{e:2lev},
possibly for longer times than $t_{max}$. On the other hand, 
the predicted evolution $\tilde\rho_\beta(t_j)$ is obtained by feeding the initial state  
$\rho_\beta(0)$ to the RNN to get the entire temporal sequence. The two evolutions are compared 
with the trace distance $D(\rho,\tilde\rho) = \frac12\Tr|\rho-\tilde\rho|$. In particular, 
we study the average trace distance $T(\rho(t),\tilde\rho(t)) = \frac1\NP\sum_\beta D(\rho_\beta(t),\tilde\rho_\beta(t))$ as a function of time.

In Fig.~\ref{fig:trace_distance} we show a numerical solution of this numerical experiment (EXP1), where we can see that,
once trained, the RNN is able to predict the evolution of a known starting state, 
up to the maximal training time $t_{max}$. On the other hand, and as expected, the accuracy of the RNN prediction rapidly deteriorates 
for $t>t_{max}$.

\subsection{Learning the master equation: a simple case}

\begin{figure}[t]
    \centering
    \includegraphics[width=.49\textwidth]{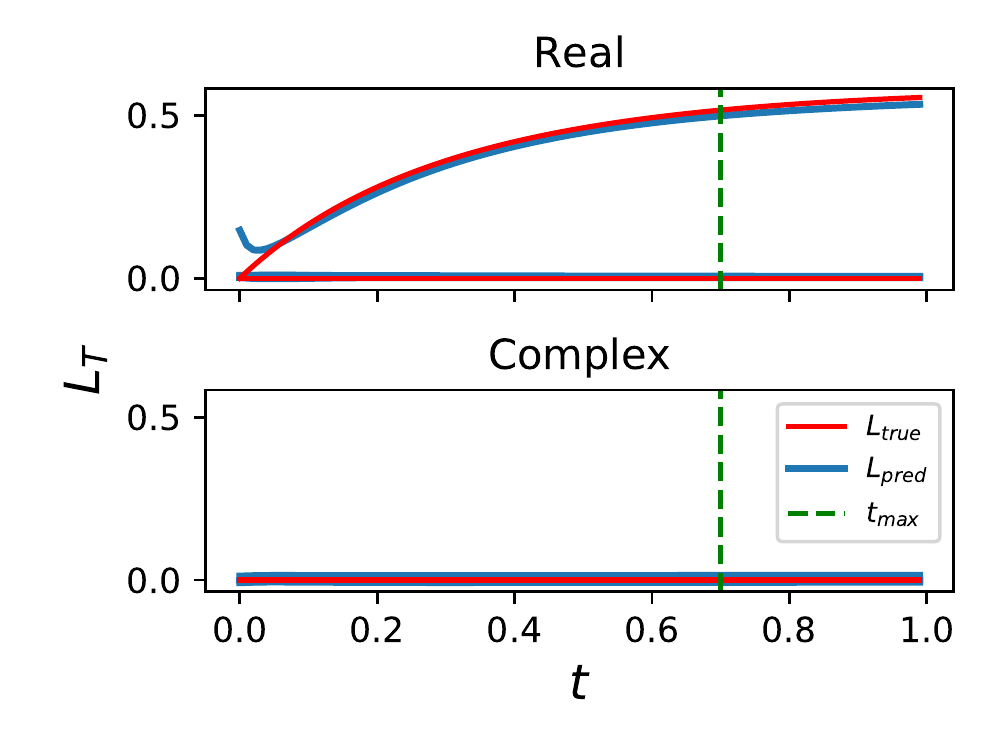}
		\caption{\textbf{(EXP2) Learning the Lindblad operator describing the
			Markovian evolution of a two level system:} the learned and known
			entries of $\frac{1}{\alpha}\sum_{\alpha}L^{\mu\equiv1}$ are compared for the
			Markovian evolution of a two-level system. We plot the time evolution of every entry of the matrix. Real and complex entries are plotted separately. 
				All the experiments run with the following parameters $\omega = 1$ for
		Eq.~\eqref{e:2lev} and $\lambda = 2$, $\gamma_0 = 0.5$ for Eq.~\eqref{e:rabi},
		and a discretization interval of $\Delta t = 0.01$. The simulations run
		over $1500$ training examples and predictions were tested over $2500$ test
		examples. All the evolutions run for a time of $t_{max} = 0.7$. The
		Frobenius norm squared was used as distance between the matrices.  
    }
    \label{fig:expA}
\end{figure}

\begin{figure}[t]
    \centering
    \includegraphics[width=.49\textwidth]{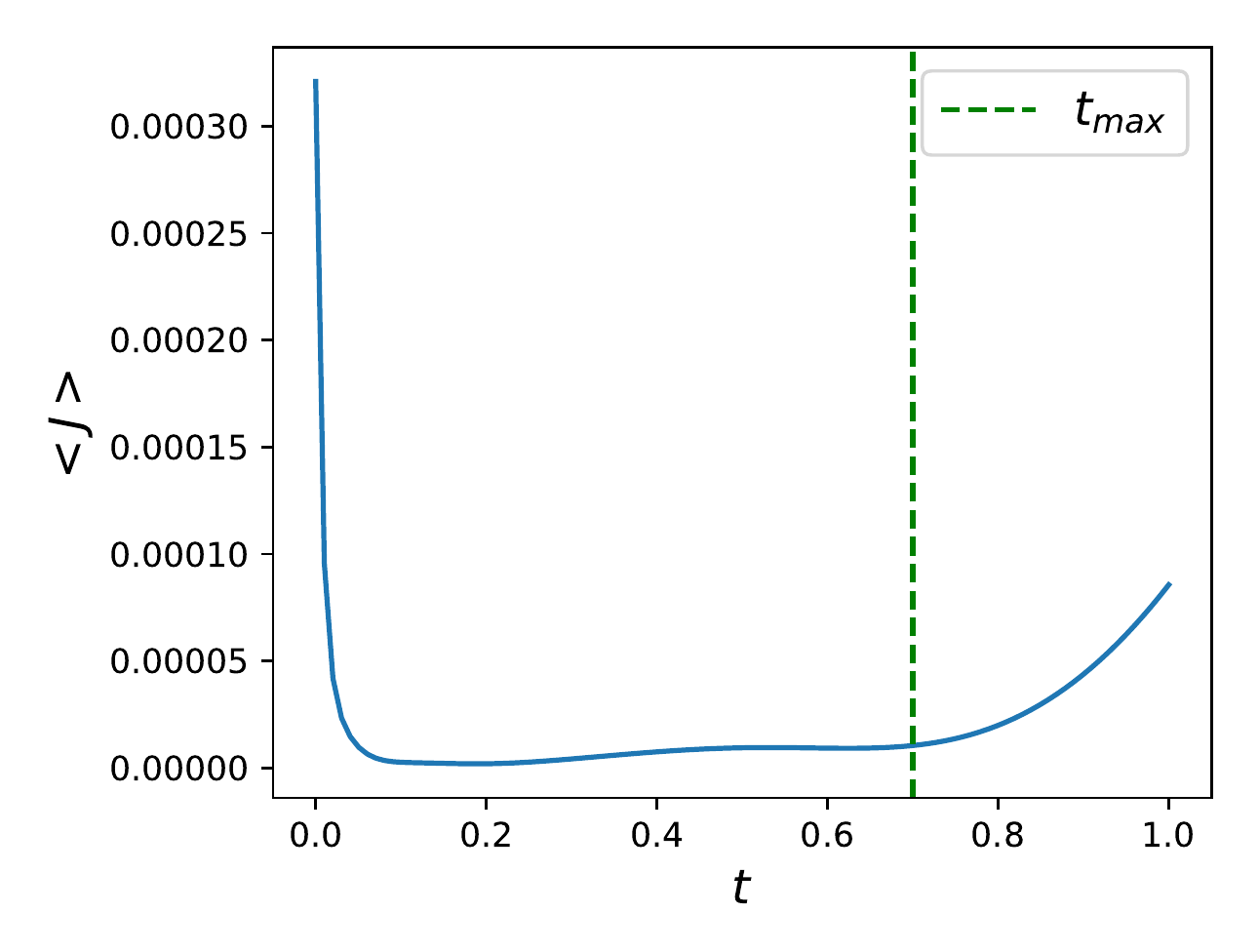}
		\caption{\textbf{(EXP2) Average cost function value \eqref{e:cost} for different times. 
		} The parameters are the same of Fig.~\eqref{fig:expA}.}
    \label{fig:expA_error}
\end{figure}

In Fig.~\ref{fig:expA} we show the solution of a second numerical experiment (EXP2) 
obtained with the same training set of the first experiment (EXP1), 
discussed in Fig.~\ref{fig:trace_distance}. While EXP1 uses a RNN to model the entire 
quantum evolution, EXP2 uses the RNN to define a QRN master equation, 
following Sec.~\ref{s:envlearning}. Training is then performed by minimising 
the QRN cost function \eqref{e:cost}, where for simplicity we assume a first-order 
Runge-Kutta integrator ($n=1$) 
and a single recurrent Lindblad operator ($\mu\equiv1$). 
In EXP2 we have chosen a Markovian regime so that the entire evolution can be modeled via 
a GKSL master equation \eqref{e:GKSL}. In this regime, the RNN Lindblad operators 
can be approximated as a simple time-dependent function, so the minimization of the 
cost function \eqref{e:cost} is equivalent to learning standard Lindblad operators. 
In Fig.~\ref{fig:expA} we compare the predicted recurrent Lindblad operators $L_\mu$ 
as a function of time, with respect to the real ones defined in Eq.~\eqref{e:2lev}. 
			In the system under study all entries of the Lindblad operators are zero (hence the flat lines in the figure) apart from one value.
As we can see, the prediction is remarkably accurate at all times, even beyond 
the training time $t>t_{max}$. The error in the prediction is shown in 
Fig.~\ref{fig:expA_error}, by plotting the cost function \eqref{e:cost} 
for different times. 
EXP2 shows that the RNN was able to learn the
evolution of the Lindblad operator $L_{t}$ for $t=0.1$ to $t_{max} =
0.7$.  In addition, the RNN was able to predict $L_{t}$ for $t>t_{max}$
which was the last time step used during training.

\subsection{Learning the master equation: non-Markovian case}\label{s:backscatt}

In this section we focus on the non-Markovian regime, where there are non-trivial 
memory effects that the RNN has to learn and reproduce. 
As in Sec.~\ref{s:statelearn}, we consider state sequences 
generated numerically by solving a master equation. 
However, unlike our previous treatment, here we consider 
a more complicated non-Markovian model of the environment, which
includes back-scattering effects. Back-scattering can refer, for instance, 
to a photon emitted to the environment  that comes back at later times. 
As such, the information transferred to the environment is not completely lost. 
To model these non-Markovian effects we consider 
two qubits evolving with the following master equation 
\begin{align}
  \label{e:backscatt}
	\frac{\partial}{\partial t}&\rho_{12}(t) =-i 
	[H_{12},\rho_{12}(t)] + \\
	&+\sum_{i=1}^2 \gamma_i(t) \left( \sigma^-_i\rho_{12}(t)\sigma^+_i
	-\frac{1}{2}\left\{\sigma^+_i\sigma^-_i,\rho_{12}(t)\right\}
\right),\nonumber
\end{align}
where $\gamma_i(t)$ has the same functional form of Eq.~\eqref{e:rabi} (but we add 
a superscript  $(i)$ to the parameters $\gamma^{(i)}_0$ and $\lambda^{(i)}$ 
that refers to qubit index) and the two-qubit Hamiltonian is 
\begin{equation}
	H_{12} = \omega \, \sigma_z \otimes I + c_1 \, \sigma_x \otimes \sigma_x + c_2 \, \sigma_y \otimes \sigma_y + c_3 \, \sigma_z \otimes \sigma_z,
	\label{e:two_q_hamilt}
\end{equation}
where $ c_1 = 0.3242$, $c_2 = 0.6723$, and $c_3 = 0.1353$.
We numerically solve Eq.~\eqref{e:backscatt} to get the data 
sequence $\rho^{\alpha}_{12}(t_j)$, where $\alpha=1,\dots,\NI$ 
indexes the different solutions obtained with different initial 
states. From these two qubit solutions we define then the state 
sequences as $\rho_\alpha(t_j)= \Tr_2[\rho^\alpha_{12}(t)]$. 
In other terms, qubit 1 is the principal system, while qubit 2 is an 
ancillary system.  Because of the coherent interaction $H_{12}$ between 
qubit 1 and qubit 2, with this approach 
we can mimic the back-action of the environment onto the system. 

\begin{figure}[t]
    \centering
    \includegraphics[width=.49\textwidth]{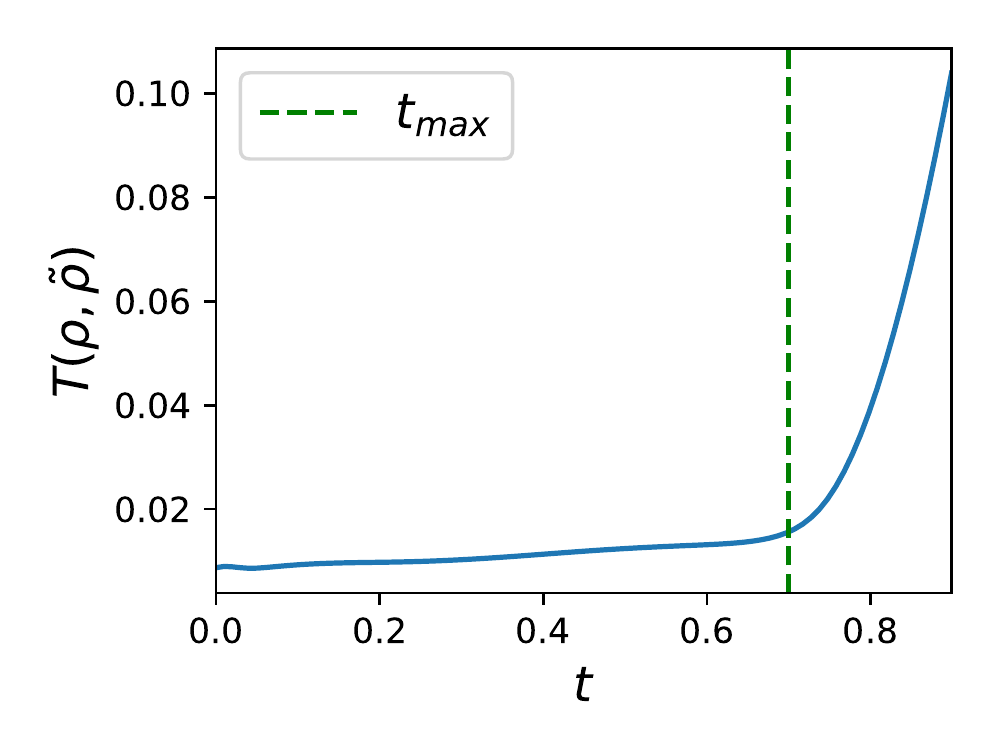}
    \caption{\textbf{(EXP3) Learning the evolution of a noisy two level non-Markovian system:} 
			the mean trace distance $T(\rho(t),\tilde\rho(t)) = \frac1\NP\sum_\beta D(\rho_\beta(t),\tilde\rho_\beta(t))$
			between true states $\rho_\beta(t_j)$ and prediction states $\tilde{\rho}_\beta(t_j)$ as a function of time. 
			The time $t_{max}$ is the maximum time used during training.
		The experiments run with the following parameters $\lambda^{(1)} = 2$, $\gamma^{(1)} _0 = 0.5$ and $\lambda^{(2)} = 1$, $\gamma^{(2)}_0 = 0.2$ for Eq.~\eqref{e:backscatt}, $\omega = 1$ for Eq.~\eqref{e:two_q_hamilt}, and a discretization interval of $\Delta t = 0.01$. The simulations run over $3000$ training examples and predictions were tested over $1000$ test examples. All the evolutions run for a time of $t_{max} = 0.7$. }
    \label{fig:trace_distance_NM}
\end{figure}
\begin{figure}[t]
    \centering
    \includegraphics[width=.49\textwidth]{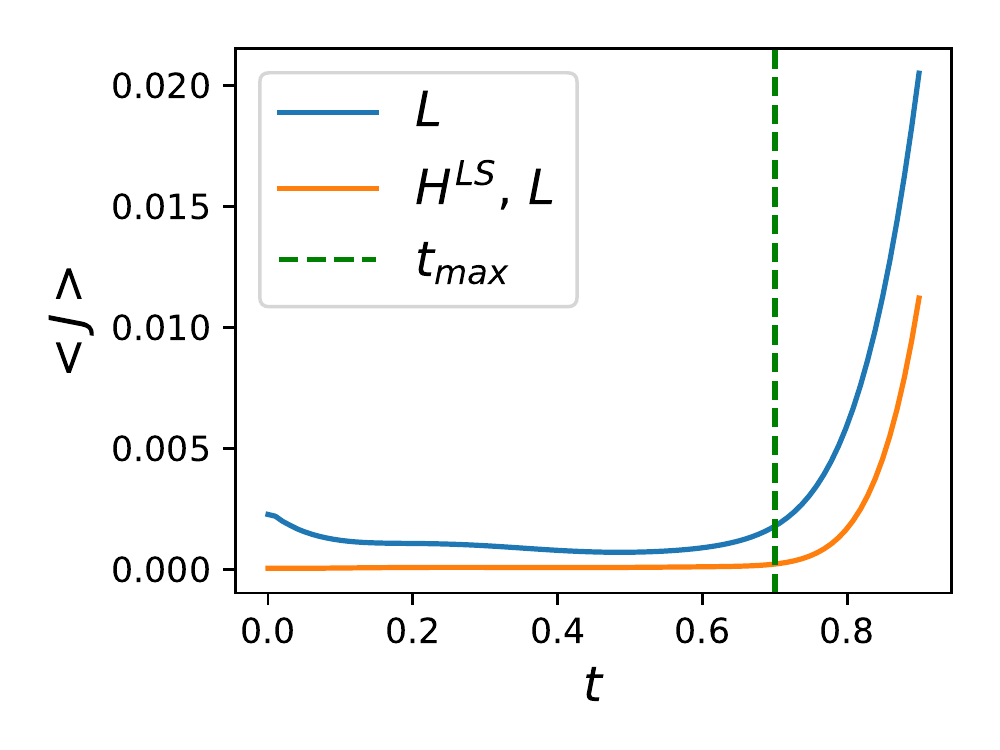}
    \caption{\textbf{(EXP4) Learning a non-Markovian memory kernel:} 
		average value of the cost function $\langle J \rangle$ on unseen examples as a function of time for a learned memory kernel with only the Lindblad Operator $L_{{\le}t}^\mu$ and with both the Lindblad Operator and the Lamb-shift Hamiltonian $H^{LS}_{{\le}t}$. The parameters are the 
		same of Fig.~\ref{fig:trace_distance_NM}. 
	The Frobenius norm squared was used as distance between the matrices.  }
    \label{fig:NM_cost}
\end{figure}

We divide our numerical results into different experiments. In EXP3, 
shown in Fig.~\ref{fig:trace_distance_NM}, we train a RNN to fully reproduce the 
entire quantum evolution, following the discussion of Sec.~\ref{s:qp}. We see
that in spite of non-Markovian effects, the resulting error is comparable to that 
of the Markovian case (EXP1) shown in Fig.~\ref{fig:trace_distance}. 

In Fig.~\ref{fig:NM_cost} we use the same training data of EXP3 to run a new experiment
(EXP4) where we train a QRN master equation, by minimising the cost function 
\eqref{e:cost}. We consider two cases: in the first one the RNN outputs  
a single recurrent Lindblad operator $L^{\mu\equiv 1}$. In the second one, 
the RNN  outputs both model both a recurrent Lindblad operator  $L^{\mu\equiv 1}$,
and the renormalized Hamiltonian, namely the Lamb-shift term $H^{LS}$. 
We note that, with a single recurrent Lindblad operator, the error is slightly larger 
than that of the Markovian case, shown in Fig.~\ref{fig:expA}. However, the resulting 
error is very low when we include also the correction Hamiltonian $H^{LS}$. 
Comparing  Fig.~\ref{fig:NM_cost} with the Markovian case, Fig.~\ref{fig:expA}, 
one can see that the error grows faster in the non-Markovian regime
after the training time $t_{\rm max}$. 
Overall, in Fig.~\ref{fig:NM_cost} we see that the RNN which learned both the recurrent 
Lindblad operator and the Lamb-shift Hamiltonian performed best, and was better able to predict the memory kernel at times greater than those seen during training.

\begin{figure}[t]
    \centering
    \includegraphics[width=.49\textwidth]{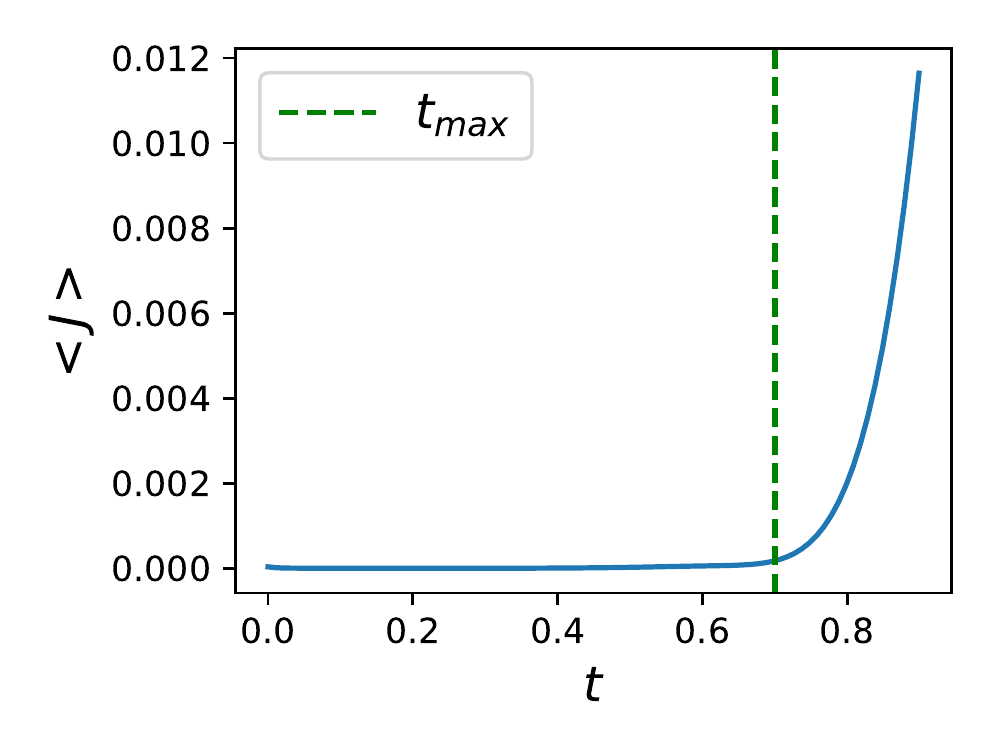}
    \caption{\textbf{(EXP5) Learning a non-Markovian master equation with different $\omega$ values:} average value of the cost function $\langle J \rangle$ on unseen examples as a function of time for a learned memory kernel with both the Lindblad operator $L_{{\le}t}^\mu$ where $\mu=2$ and the Lamb-shift Hamiltonian $H^{LS}_{{\le}t}$. 
			$t_{max}$ specifies the maximum time in the range of times used for training. 
    All the experiments run with the following parameters  $\lambda^{(1)} = 2$, $\gamma^{(1)}_0 = 0.5$ and $\lambda^{(2)} = 1$, $\gamma^{(2)}_0 = 0.2$ for Eq.~\eqref{e:backscatt}, $\omega$ uniformly sampled in the interval $[0.5,1.5]$ for Eq.~\eqref{e:two_q_hamilt}, and a discretization interval of $\Delta t = 0.01$. The simulations run over $3000$ training examples and predictions were tested over $1000$ test examples. All the evolutions run for a time of $t_{max} = 0.7$. The Frobenius norm squared was used as distance between the matrices.  }
    \label{fig:NM_cost_multi_Omega1}
\end{figure}

Finally, in Fig.~\ref{fig:NM_cost_multi_Omega1} we run a different numerical experiment 
(EXP5). In EXP5 the training set is composed by data sequences where 
the qubit frequency $\omega$ in Eq.~\eqref{e:two_q_hamilt} is not fixed, but rather 
uniformly sampled from 0.5 to 1.5. The sampled frequency is used as an extra input to the RNN.
This corresponds to the experimentally relevant case where
the qubit frequency can be externally tuned to a known value. Uncertainty about this frequency can be estimated from the correction Hamiltonian $H^{LS}$, which is learned from the data. 
In Fig.~\ref{fig:NM_cost_multi_Omega1} we see that the error in EXP5 is remarkably low. 
Based on the success of this numerical experiment, we propose the following general strategy to introduce prior knowledge about the system. We can define a RNN where the known properties of the system, e.g. its Hamiltonian, are added as an extra input.  Training is performed using datasets of quantum state sequences and their respective Hamiltonian, where the latter is sampled from the space of experimentally relevant Hamiltonians. 
The remarkable accuracy shown Fig.~\ref{fig:NM_cost_multi_Omega1} suggests that this procedure  forces the RNNs to better explore the manifold of quantum state sequences, to produce a more accurate and robust prediction.

\section{Conclusions and Perspectives}\label{s:conclusions}

We have studied quantum state evolution using RNNs. 
We have shown that even when the system is interacting with a complicated 
surrounding environment, RNNs offer an accurate and robust tool for 
modeling and reconstructing the quantum evolution, both in the Markovian and 
non-Markovian regime, and even when there are back-scattering effects
from the environment. 

We have introduced two approaches for modelling open quantum systems with RNNs. In 
the first one, a deep RNN is trained to learn the entire quantum evolution,  
namely to reproduce the time sequence $\rho(t_j)$ given an initial state $\rho(0)$. 
In the second approach, we use a deep RNN to define a non-Markovian master equation 
where the memory kernel takes a convenient mathematical form, namely that 
of the GKSL equation. 
In our master equation the non-Markovian memory effects are taken into account 
by the structure of the RNN cells. 
The observed success of our approaches stems from the ability of 
RNNs to learn temporal sequences, where the future depends on the entire past. 
Our RNN-based master equation can be used as a convenient mathematical framework
to write non-Markovian memory kernels without resorting to perturbative treatments, and
to learn their unknown parameters directly from the data.

Many extensions of our work are possible. Many-body systems with exponentially large 
Hilbert spaces could be considered by using RNNs which output a compressed representation
of the state, such as tensor networks \cite{cui2015variational,vidal2008class}, 
restricted Boltzmann machines~\cite{torlai2018neural} or  variational
autoencoders~\cite{rocchetto2017learning}.  
The predictions after the training time can be improved by considering higher order
Runge Kutta integrators and by modifying the structure of the RNN cell to have 
a fixed point in the infinite time limit, as expected from physical principles. 
Although RNNs proved to be a
powerful tool for modelling open quantum systems, the machine learning
literature offers a number of other methods, such as Kalman Filters or models
with Gaussian Process transitions, that appear particularly promising. For
example, Gaussian Process State Space Models~\cite{frigola2014variational}
allow one to model prior information on the system and return
Bayesian estimates of the uncertainties. Both these features are desirable in a
quantum context:  prior knowledge on the system, such as the form of the
noise-free Hamiltonian, may be used to further reduce the complexity of the
model, while approximate values for the uncertainties might enable better
control of experimental inaccuracies. 
It would be interesting to define quantum master equations based on these tools 
and to check their performances against our QRN master equation.

Finally, in more physical terms, it would be interesting to study what happens when 
{\it partial} information about the system is available. An experimentally relevant 
case is when the initial state $\rho(0)$ is known, but one has access to a limited 
set of expectation values $\langle{ A_k }\rangle_{\rho(t_j)}$, where the observables 
$A_k$ are not enough to tomographically reconstruct 
the states $\rho(t_j)$. This possibility may be considered, using our framework,
by introducing a cost function 
between expectation values, rather than between density operators. A further challenge is 
then to model the disturbance of the measurement onto the system, namely the wave function
collapse. This can be done using the process tensor formalism \cite{pollock2018non},
which provides an avenue for generalising our approach in the presence of feedback. 
It would be interesting to study the performance of RNNs to model these 
experimentally relevant quantum evolutions.

\begin{acknowledgements}
The authors would like to thank G. Carleo, A.D. Ialongo, and M. Paternostro for valuable discussions. 
L.B. is supported by the UK EPSRC grant EP/K034480/1. 
E.G. is supported by the UK EPSRC grant EP/P510270/1.
A.R. is supported by an EPSRC DTP Scholarship and by QinetiQ.  S.S. is supported by the Royal Society, EPSRC, the National Natural Science
Foundation of China, and the grant ARO-MURI W911NF17-1-0304 (US DOD, UK MOD and UK EPSRC under the Multidisciplinary University Research Initiative). 
We gratefully acknowledge the support of NVIDIA Corporation with the donation of the
Titan Xp GPU used for this research.
Part of this work was conducted while A.R. and S.S. were at the Institut Henri Poincar\'e in Paris. 
\end{acknowledgements}

\appendix 
\section{ Gated Recurrent Unit} \label{a:GRU}
The GRU state $s_{t}$ is a linear interpolation of the previous state $s_{t-1}$ and a candidate state $\tilde{s}_{t}$. The candidate state is a function of the cell input $x_{t}$ and the previous state given by:
 \begin{align}
 \begin{split}
	 z_{t}&=\sigma(W_{z}x_{t}+U_{z}s_{t-1}+b_{z})~,\\
	 r_{t}&=\sigma(W_{r}x_{t}+U_{r}s_{t-1}+b_{r})~,\\
 \tilde{s}_{t}&=\tanh(Wx_{t}+U(r_{t}\circ s_{t-1})+b)~,\\
 s_{t}&=(1-z_{t})\circ s_{t-1}+z_{t}\circ  \tilde{s}_{t}~,
 \end{split}
 \label{GRU}
 \end{align}
where $\circ$ denotes the entry-wise (Hadamard) product, $\sigma$ denotes the sigmoid function, $x_{t}$ is the input to the cell and $W$, $U$ and $b$ are trainable parameters, which do not explicitly depend on $t$. GRU cells can be stacked to form a deep GRU network. The input $x^{j}_{t}$ for a depth $j$ cell at time $t$ is the state from the cell in the previous layer $s^{j-1}_{t}$.

The GRU state $s_{t}$ can be thought of as a mixture of the previous state $s_{t-1}$ and the candidate state $\tilde{s}_{t}$. The weighting of the mixture is controlled by $z_{t}$ which is known as the \textit{update gate}. We can see that if the entries of $z_{t}$ are zero then the candidate state is ignored and the previous state becomes the new state. 

The candidate state $\tilde{s}_{t}$ is a function of the previous state $s_{t-1}$ and the current input $x_{t}$. The relative contribution of the previous state to the candidate state is controlled by $r_{t}$ which is known as the \textit{reset gate}.
 
A more detailed discussion of GRU cells can be found in Ref.~\cite{cho2014learning}.

Note that the output of the network is a real vector. In order to encode complex matrices into the RNN we use the following procedure. Assume that we want to encode a matrix $M \in \mathbb{C}^{m\times m}$. $M$ can be decomposed into a real and imaginary part $M = M_{\mathrm{Re}} + i M_{\mathrm{Im}}$. We require the output of the RNN to be a vector $o\in \mathbb{R}^{2m^2}$ such that the first $m^2$ elements encode in row-major order the entries of $M_{\mathrm{Re}}$ and the second $m^2$ elements encode in row-major order the entries of  $M_{\mathrm{Im}}$.

All GRU networks used in this work consisted of $2$ GRU layers with output dimension = $40$, followed by $1$ fully connected layer. The Adam optimizer was used for training with initial learning rate $0.01$~\cite{kingma2014adam} and batches of $32$ examples. Training was stopped after each example was seen $60$ times. 
In EXP2 the fully connected weights were regularized using weight decay \cite{krogh1992simple}, with the L2 penalty factor set to 0.001.
The networks were implemented using the Keras $2.2.0$ and Tensorflow $1.8$ frameworks~\cite{chollet2015keras,abadi2016tensorflow}. Network weights were initialized using the Keras defaults.

\end{document}